\documentclass[twocolumn,a4paper,superscriptaddress,aps,prb]{revtex4-1}
\pdfoutput=1

\usepackage{amsmath}
\usepackage{graphicx}
\usepackage[english]{babel}
\usepackage[version=4]{mhchem}
\usepackage[breaklinks]{hyperref}
\hypersetup{
	pdfnewwindow=true,      % links in new window
    colorlinks=true,       % false: boxed links; true: colored links
    linkcolor=blue,          % color of internal links
    citecolor=blue,        % color of links to bibliography
    filecolor=blue,      % color of file links
    urlcolor=blue           % color of external links
}

\begin{document}
\title{Spatial and spectral dynamics in STEM hyperspectral imaging using random scan patterns}

\author{Alberto Zobelli}
\email{alberto.zobelli@u-psud.fr}
\affiliation{Laboratoire de Physique des Solides, Univ. Paris-Sud, CNRS UMR 
	8502, F-91405, Orsay, France}

\author{Steffi Y. Woo}
\affiliation{Laboratoire de Physique des Solides, Univ. Paris-Sud, CNRS UMR
	8502, F-91405, Orsay, France}

\author{Luiz H. G. Tizei}
\affiliation{Laboratoire de Physique des Solides, Univ. Paris-Sud, CNRS UMR
	8502, F-91405, Orsay, France}

\author{Nathalie Brun}
\affiliation{Laboratoire de Physique des Solides, Univ. Paris-Sud, CNRS UMR
	8502, F-91405, Orsay, France}

\author{Anna Tararan}
\affiliation{Laboratoire de Physique des Solides, Univ. Paris-Sud, CNRS UMR
8502, F-91405, Orsay, France}

\author{Xiaoyan Li}
\affiliation{Laboratoire de Physique des Solides, Univ. Paris-Sud, CNRS UMR
	8502, F-91405, Orsay, France}

\author{Odile St\'ephan}
\affiliation{Laboratoire de Physique des Solides, Univ. Paris-Sud, CNRS UMR
	8502, F-91405, Orsay, France}

\author{Mathieu Kociak}
\affiliation{Laboratoire de Physique des Solides, Univ. Paris-Sud, CNRS UMR
8502, F-91405, Orsay, France}

\author{Marcel Tenc\'e}
\email{marcel.tence@u-psud.fr}
\affiliation{Laboratoire de Physique des Solides, Univ. Paris-Sud, CNRS UMR
8502, F-91405, Orsay, France}

\begin{abstract}
The evolution of the scanning modules for scanning transmission electron microscopes (STEM) has realized the possibility to generate arbitrary scan pathways, an approach currently explored to improve acquisition speed and to reduce electron dose effects. In this work, we present the implementation of a random scan operating mode in STEM achieved at the hardware level via a custom scan control module. A pre-defined pattern with fully shuffled raster order is used to sample the entire region of interest. Subsampled random sparse images can then be extracted at successive time frames, to which suitable image reconstruction techniques can be applied. With respect to the conventional raster scan mode, this method permits to limit dose accumulation effects, but also to decouple the spatial and temporal information in hyperspectral images. 
We provide some proofs of concept of the flexibility of the random scan operating mode, presenting examples of its applications in different spectro-microscopy contexts: atomically-resolved elemental maps with electron energy loss spectroscopy and nanoscale-cathodoluminescence spectrum images. By employing adapted post-processing tools, it is demonstrated that the method allows to precisely track and correct for sample instabilities and to follow spectral diffusion with a high spatial localization.

\end{abstract}

\maketitle

\section{Introduction}
In the last twenty years, significant improvements on the spatial resolution of scanning transmission electron microscopes (STEM) have been achieved by the widespread implementation of spherical aberration correctors. Currently, atomically-resolved images are routinely obtained for a wide range of materials going from semiconductors, nanostructures, to functional oxides, etc. The structural information can be complemented by the synchronous acquisition of several spectroscopic signals accessible, such as electron energy loss, X-ray emission or cathodoluminescence (CL), which can then be bundled into hyperspectral images.

Besides the instrumental characteristics, the ultimate spatial resolution obtainable in spectro-microscopy is strongly limited by the radiation sensitivity of the sample. Indeed, higher spatial resolution corresponds to a smaller illumination area and number of scattering centers, which imply higher electron doses for a given signal intensity collected. Irradiation effects have a much stronger impact in hyperspectral imaging since inelastic scattering processes have significantly lower cross sections than elastic scattering employed in STEM imaging, thus hindering the signal collection efficiency.
Furthermore, within the finite time required for recording a hyperspectral image, the sample might drift, transform and/or the spectroscopic signals of interest evolve.
Therefore, a hyperspectral image does not always portray the sample in a definitive state since different pixels are acquired at different times. These unavoidable drawbacks motivate the development of novel operating modes that could reduce sample dependent limitations in STEM spectro-microscopy. 
 
The use of unconventional scan pathways is a promising route currently explored to improve acquisition speed, reduce electron dose\cite{stevens2018sub} and, in certain cases, to avoid distortions associated with scan\cite{sang2016dynamic} and sample instabilities. In a STEM, the electron beam is typically scanned over the region of interest following a sawtooth-path to fill all pixels of a rectangular frame. Alternative spiral scans have also been recently proposed for reducing fly-back image distortions arising at high scanning rates.\cite{sang2016dynamic,sang2017precision,li_dyck_kalinin_jesse_2018} In the last years, subsampling has been extensively discussed as a very effective strategy for dose-reduced image acquisition. Indeed, STEM images are often associated with a certain degree of over-redundancy, and a good approximation of the full image can already be obtained from an appropriate subset of pixels. The subsequent application of image reconstruction techniques can permit to fill the missing data within the image matrix.\cite{stevens2013potential,trampert2018should} The direct acquisition of sparse images presents the important advantage of reducing radiation dose, acquisition time and data size.\cite{binev2012compressed,trampert2018should} An easy and implementable choice for sparse sampling is a fully random distribution of the pixels, which permit high subsampling with minimal reconstruction distortions.\cite{Lustig2008} The applicability of different reconstruction algorithms for S(T)EM imaging has been demonstrated on simulated random scan images obtained by extracting a subset of randomly chosen pixels from full images.\cite{binev2012compressed,stevens2013potential,anderson2013sparse}
The effectiveness of these methods has also been shown for multi-dimensional data structures such as tomography image series,\cite{saghi2015reduced,donati2017compressed} STEM-ptychography,\cite{stevens2018ptycho}, energy-dispersive X-ray spectroscopy,\cite{hujsak2018} and electron energy loss spectroscopy (EELS) hyperspectral images.\cite{monier2018reconstruction}%,monier2019spimBPFA}
The direct acquisition of random sparse STEM images can be attained by scripting the acquisition control software,\cite{hwang2017towards} but this method suffers from a very slow response time. More efficiently, B\'ech\'e and coworkers developed an experimental set up which integrates a fast electromagnetic shutter: the electron probe follows the conventional scan pattern but the beam can be turned off at randomly selected pixels.\cite{beche2016development} Whereas the method permits a reduction of the electron dose, the acquisition time remains invariant with respect to a standard scan. Other methods that generate sparse random or pseudo-random scans have been proposed by exploiting the position uncertainties due to the deflection coils dynamics.\cite{anderson2013sparse,kovarik2016implementing} 
The interest of sparse sampling and compressive sensing has been stressed for TEM and STEM imaging but only minor attention has been dedicated to hyperspectral imaging, although irradiation effects, acquisition times and data sizes are significantly more important in this context. 

In this work, we present an original implementation of an effective random scan acquisition mode in STEM obtained directly using the scanning control unit. The setup permits an extended control of the scanning parameters and therefore offers a large flexibility on the acquired image data structure. The complete image matrix is filled in a fully randomized fashion, each pixel being stamped with its acquisition time.  A series of subsampled random sparse images can then be extracted at successive time frames and inpainting image reconstruction applied. This method permits to decouple the space and time information and to monitor the sample and spectral signals dynamics. After illustrating the operating principle (section \ref{operating-principle}) and its practical implementation (section \ref{practical-implementation}), we provide examples of EELS and nano-cathodoluminescence (nano-CL) hyperspectral images acquired with the newly implemented random scan mode (section \ref{proof-of-concept}). With respect to the conventional raster scan mode, we demonstrate that random scan permits to limit dose accumulation effects, but we also emphasize the possibility of tracking and correcting for sample drift and to monitor the spectroscopic signal dynamics over time in hyperspectral images.

\section{Random scan operating principle}\label{operating-principle}

\begin{figure*}[tb]
	\centering
	\includegraphics[width=0.85\textwidth]{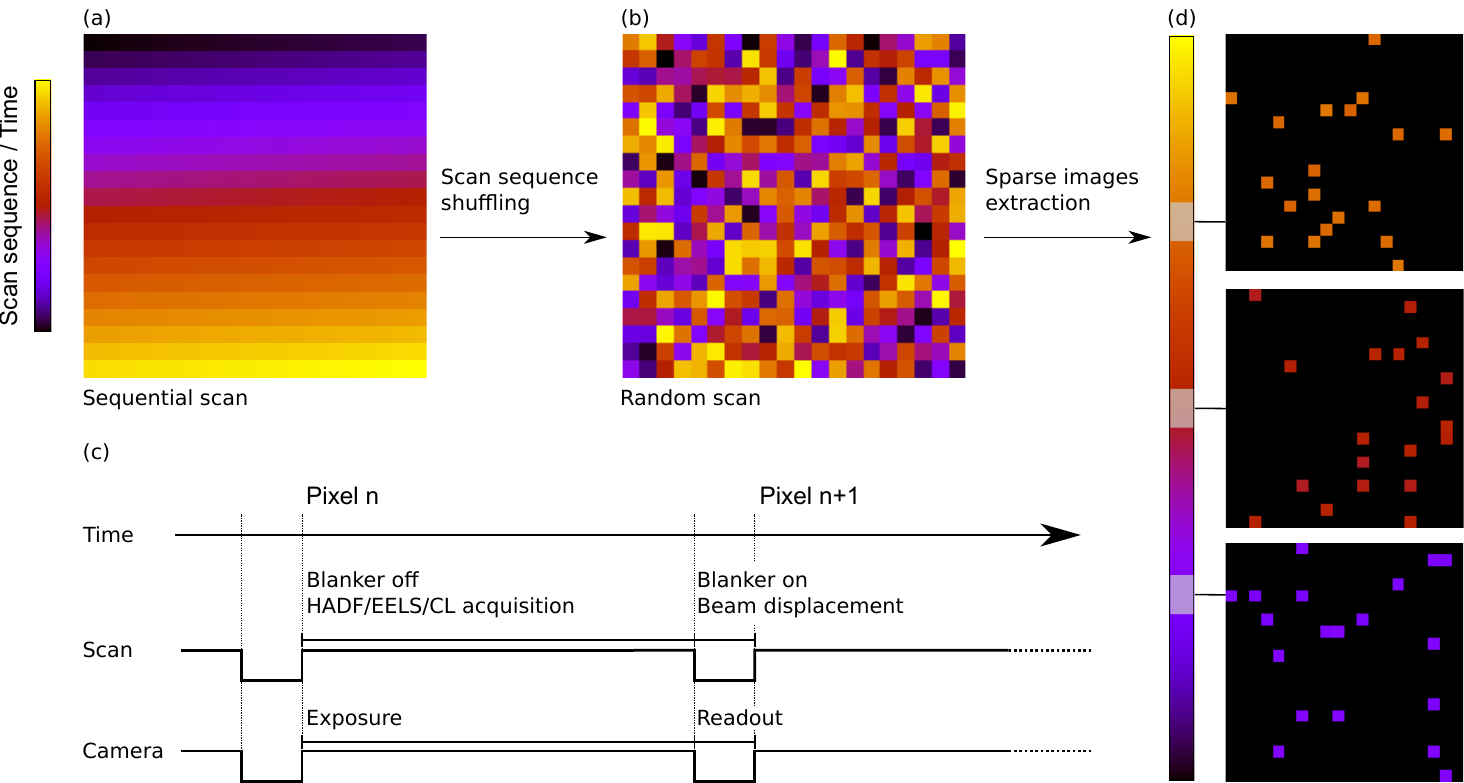}
	\caption{Principle of the standard raster scan (a) and the random scan sequence (b). The color scale represents the pixel acquisition order. (c) Schematic representation of the synchronization between illumination, beam displacement, blanking, camera exposure, and read-out. (d) Extraction of sparse random images at given time frames.		
	 \label{random-scan}}
\end{figure*}

In Fig. \ref{random-scan}(a) is represented the sawtooth-scan order typically employed in STEM hyperspectral imaging: the region of interest is scanned from left-to-right and top-to-bottom. In the proposed random-scan mode, each pixel is visited only once but in an aleatory order. In the first step, the list of pixel coordinates of the scanning matrix is built. This list is then shuffled to define the order sequence for pixel illumination; the matrix representing the pixel scanning order can be stored for post-processing [Fig. \ref{random-scan}(b)]. In order to avoid spurious image distortions, acquisition and displacement should be performed in separated time frames [Fig. \ref{random-scan}(c)].

An advantage of the random-scan mode is the reduction of dose accumulation. After illumination, the system transitions to its ground state within a characteristic time defined by diffusion and recombination processes (charge or thermal diffusion, radiative recombination, etc.). The illuminated area might therefore interact with previously visited regions or even overlap with an already excited state since excitation volumes can be larger than the probe size.
This dose accumulation effect can lead to image artifacts or induce radiation damages.\cite{egerton2019radiation} In a traditional sequential scan, successively acquired pixels are adjacent and therefore accumulation effects are maximized. A typical example is charge artifacts observed in scanning electron microscopy (SEM) and secondary ion mass spectrometry (SIMS) imaging of insulating samples. 
A way to minimize this issue is to ensure that the illuminated area is far enough from a region exposed short time before. The random scan mode grants that this criteria is satisfied for the large majority of the image pixels. An alternative method consists of scanning the region of interest with several passes following a standard sequential pattern, and acquiring only a fraction of the image pixels per pass to fill in the full image. This method, already implemented in SEM imaging,\cite{thong2001reduction} has the drawback of significantly elongating the total acquisition time while only a single pass is required in the random scan mode.

Reducing irradiation effects can also be achieved by acquiring sparse images containing only a fraction of well-interspersed pixels of the frame. The full image can then be reconstructed using appropriate inpainting techniques.\cite{stevens2013potential,anderson2013sparse}
The effectiveness and applicability of such image reconstruction methods for STEM-HAADF imaging has been extensively discussed in Ref. \citenum{binev2012compressed}.

Since the sparsest solution which is consistent with the target image is \textit{a priori} unknown, the choice of a predefined sampling rate might be responsible for a significant loss of information.\cite{binev2012compressed}
An important advantage of employing an effective random scan acquisition mode arises from the fact that any subset of pixels is spanned homogeneously all over the region of interest. This permits to sequentially build sparse images with a progressively increasing sampling density, thus image reconstruction techniques can then be applied in real-time to gradually improve the target image resolution. Moreover, the acquisition can be interrupted once the required level of accuracy is achieved. The method allows therefore to operate at the effective low-dose limit compatible with the signal of interest. Furthermore, it permits fast surveys with a wide field of view for effective tracking of regions of interest, which can be imaged successively at a higher sampling rate.

In a traditional sequential raster scan, the space and time information are coupled and the temporal evolution of the imaged region or of the spectroscopic signals during the acquisition cannot be recovered. This limitation can be overcome by operating in a random scan mode, and time-stamping the pixels. The acquired data can be represented as a random sparse matrix having two space and one time coordinate for simple images, and an additional spectroscopic dimension for hyperspectral imaging. 
This object can be segmented into successive time windows and a series of random sparse images/hyperspectral images can be built [Fig. \ref{random-scan}(d)]. Each time-frame can then be used for reconstructing the full target image. The recovery of the time dependence of signals has several important applications: it allows to track the objects' motion and, in pertinent cases, to correct for image drifts; it permits to monitor the signal dynamics in hyperspectral images; it permits to identify changes on images or spectra which can be associated to irreversible transformations of the sample due to irradiation.
Examples of the application of time-slicing of hyperspectral images acquired in the random scan mode will be presented in section \ref{proof-of-concept}.

\section{Practical implementation in STEM spectro-microscopy}\label{practical-implementation}

The evolution of the scanning modules for scanning transmission electron microscopes (STEM) now allows to generate arbitrary scanning patterns and to explore new illumination modes. Through the development of a custom scan control module, we have implemented the random-scan method described here on two dedicated STEM microscopes: a VG-HB501 and a spherical aberration-corrected Nion UltraSTEM 200. After defining the image matrix, the pixel acquisition order is shuffled at the start and the so-defined random scan path is loaded into a dedicated scan generator. For each pixel, the beam illuminates the specimen only for the time required for the signals acquisition while the camera is in the exposure mode [Fig. \ref{random-scan}(c)]. A fast electromagnetic blanking system with typical blanking speed below 100 ns diverts the beam from the specimen during the beam displacement from pixel to pixel, meanwhile the camera is in an off/readout mode. This avoids image distortion and spurious acquisitions due to the finite dynamical response of the beam displacement coils. 

The time response of the current in the deflecting coils implies that the nominal beam position can only be asymptotically achieved, and therefore the displacement time has to be adapted to a required level of accuracy. A typical criterion is to assume that the illuminated spot position should deviate from its nominal value by less than half of the pixel size. In the random scan mode, the longest displacement for successive pixels occurs for a movement between the extremes of a line or a column (vertical and horizontal movements are decoupled due to the independent circuits controlling the displacements along the two axis). The accuracy condition is achieved when the current attains $(N-1)/N$ times its target value, where $N$ is the number of pixels on a row. The displacement time ($t$) should then satisfy the condition $ t > \tau \ln N$, where $\tau$ is the time constant of the circuit, typically of the order of some tens of $\mu$s. This implies that displacements in random-scan mode take about one--two orders of magnitude longer than the usual dwell time for STEM imaging (about few $\mu$s), even if shorter times could be obtained by the use of low-inductance magnetic coils or electrostatic deflectors. On the basis of these arguments, Kovarik \textit{et al.}\cite{kovarik2016implementing} excluded that a pure random-scan method would be viable for STEM imaging, and proposed an alternative line-hopping approach to scan randomization for optimizing the acquisition speed. However, typical illumination times in hyperspectral imaging are much longer, of the order of tens of ms, and therefore the displacement time in the random scan approach would not dominate the total acquisition time. In our implementation, the beam is displaced and blanked during the camera readout time, typically 1 ms, and therefore there is negligible time difference for acquiring hyperspectral images in the standard sequential scan and in the new random scan mode.

\begin{figure*}[tb]
	\centering
	\includegraphics[width=\textwidth]{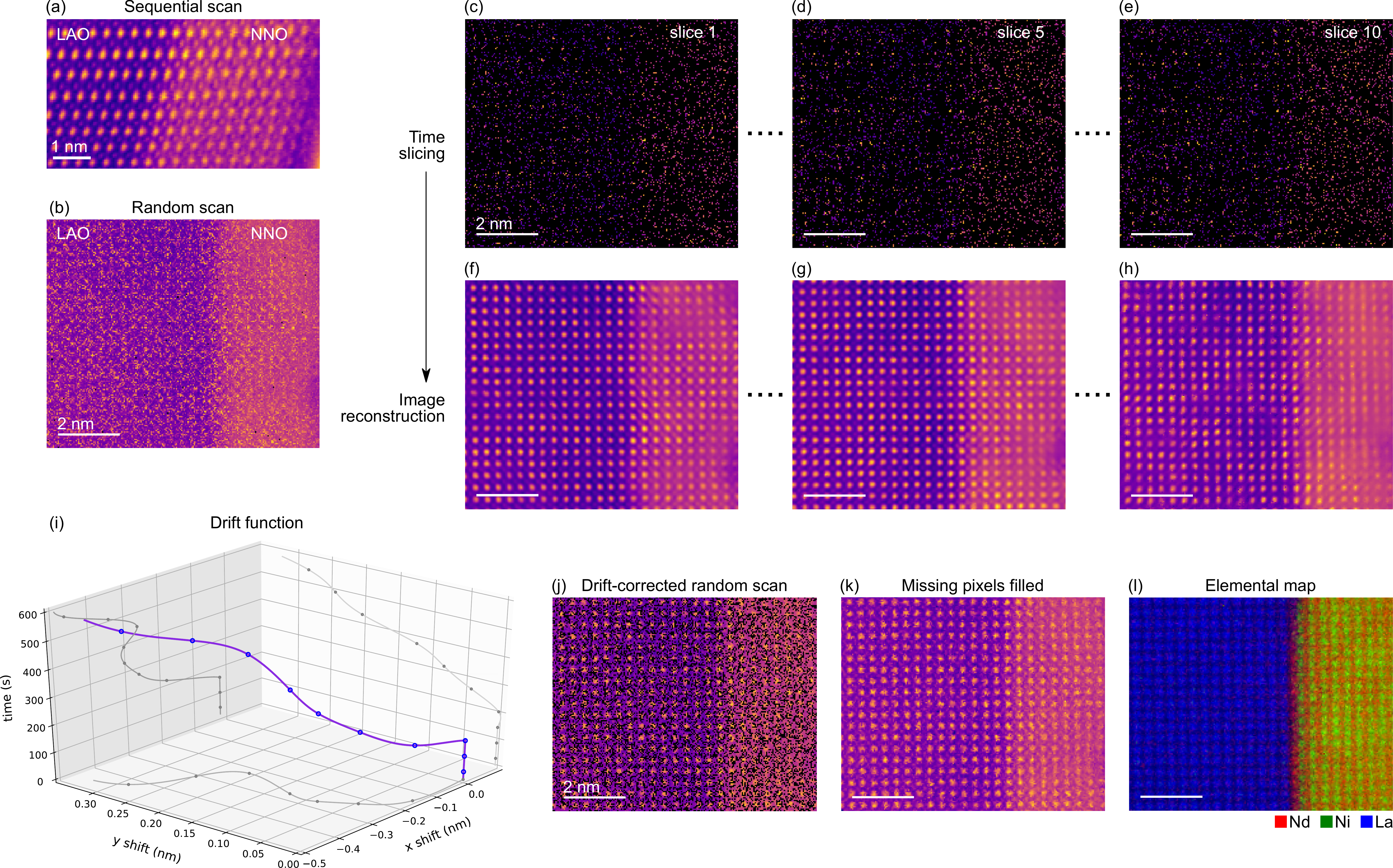}
	\caption{STEM-HAADF image of the \ce{NdNiO3-LaAlO3} (NNO--LAO) interface acquired (a) in the standard sequential raster scan mode and (b) in the random scan mode. (c--e) Time slices (10\% of total pixels) derived from the random scan image, and (f--h) corresponding images reconstructed using the BPFA method. (i) Drift function as derived from the reconstructed image stack. (j) Drift-corrected random-scan HAADF image, and (k) the same image where missing pixels have been filled using a cubic interpolation. (l) Atomically-resolved elemental maps obtained from the core-loss EELS hysperspectral image acquired simultaneously with the random-scan HAADF image. Maps extracted from the Nd-$M_\text{4,5}$, Ni-$L_\text{2,3}$, and La-$M_\text{4,5}$ edges are represented in red, green and blue, respectively. \label{drift-correction}}
\end{figure*}

\section{Proofs of concept}\label{proof-of-concept}

The random scan method can be very generally employed in hyperspectral imaging with the sole limitation of longer probe displacements with respect to a standard raster scan. As already discussed, its potential interest arises in the spectroscopy of systems sensitive to electron irradiation, and in monitoring the spectral signal and image dynamics. Here we present some proofs of concept of the flexibility of the random scan method illustrating its applications in different spectro-microscopy contexts: atomically-resolved EELS mapping and nano-CL spectrum images.

\subsection{Tracking and correcting for sample instabilities in STEM hyperspectral imaging}

Specimen drifts are very often encountered in STEM and are usually caused by limited mechanical stability of the sample holder, sample thermalization or Coulomb repulsion consequent of charge accumulation. 
Drift is, for instance, very difficult to fully eliminate in cryo-STEM, as the sample holder is cooled to liquid nitrogen temperatures and thus experiences much larger thermal gradients.\cite{savitzky2018image} Instability-induced image deformations can strongly deteriorate the imaging capabilities of modern STEM microscopes which can reach sub-\r{AA}ngstrom resolution and sub-picometer precision. Distortions are even more pronounced in hyperspectral imaging compared to HAADF imaging due to the longer dwell times. Whenever the drift origin cannot be eliminated, images could be corrected post-acquisition by displacing individual pixels on the basis of a time-drift function. However, this information can be only poorly estimated by adopting strong assumptions on the structure under observation.

As an example of such problems, Fig. \ref{drift-correction}(a) presents the high-angle annular dark-field (HAADF) image synchronously acquired with an EELS-hyperspectrum image of an atomically-sharp \ce{NdNiO3-LaAlO3} (NNO-LAO) interface\cite{Preziosi-18} performed on a Nion UltraSTEM 200 microscope using the standard sequential scan. Both materials constituting the heterostructure are insulating at room temperature, hence sample drift and the subsequent strong image deformation observed arise most probably from charge accumulation in previously scanned pixels. Whereas image rows can be individually realigned post-acquisition by assuming a straight interface, deformation parallel to the interface cannot be corrected as such. This imply that, although the imaging system is capable of a very high spatial resolution, the utmost performance cannot be reached: individual atomic columns can be visualized but image distortions hinder any precise quantitative measurement. 

Ultimate imaging capabilities can be recovered by combining the random scan acquisition mode described with a post-treatment workflow designed to segment, cross-correlate, and realign sparse image stacks. Fig. \ref{drift-correction}(b) presents the HAADF/EELS-hyperspectral image acquired using similar illumination conditions as Fig. \ref{drift-correction}(a), but in random scan mode. Now adjacent pixels are not consecutively acquired and therefore, the resulting image does not appear distorted but instead heavily blurred with a loss of atomic resolution. However, the use of time slicing combined with image reconstruction allows for a very effective and general way to measure and correct sample drifts and to recover the atomic resolution. Fig. \ref{drift-correction}(c--e) presents random sparse images slices obtained by segmenting the complete random-scanned-HAADF image in 10 successive time windows.
Each slice can be considered as a capture of the sample motion within a given time interval.
Then, target images are reconstructed by applying the beta-process factor-analysis (BPFA) algorithm,\cite{paisley2009nonparametric, zhou2012nonparametric} on which a dictionary of functions is learned from the sparse images and used to build the data under study by patches. Image inpainting using BPFA has been proven to provide high quality reconstruction for high-resolution STEM images using very limited pixel subsets.\cite{stevens2013potential} In this case, individual atomic columns can be recovered from extracted subsets having only 10\% of the total pixels of the complete image [Fig. \ref{drift-correction}(f--h)]. Relative image shifts have been measured by using the image correlation routines commonly employed for the realignment of TEM image series,\cite{guizar2008efficient} and the derived drift-time function has been represented in Fig. \ref{drift-correction}(i), showing an average drift speed of about 1 \AA/min. This information can then be used for correcting the acquired image by realigning and successively averaging the segmented STEM-HAADF images or by applying the estimated shift to each pixel individually. The obtained drift-corrected image is presented in Fig. \ref{drift-correction}(j). The reconstructed images might not present an homogeneous sampling: because of the drift, realigned pixels might be denser or sparser in subregions or along specific directions and unsampled regions might be evident. The visual rendering has been improved by filling the missing pixels (34\% of the total pixels) using an interpolation routine, as shown in Fig. \ref{drift-correction}(k). 

The recovered image presents a very straight interface and an undistorted lattice. The same drift correction can then be applied to time slices extracted from the full core-loss EELS hyperspectral image synchronously acquired with the STEM-HAADF image. Atomically-resolved elemental maps can then be extracted from the realigned hyperspectral image as shown in Fig. \ref{drift-correction}(i), where intensity maps of Nd-$M_\text{4,5}$, Ni-$L_\text{2,3}$, and La-$M_\text{4,5}$ edges are represented in red, green and blue, respectively. The aforementioned results have been obtained using a drift-correction routine which does not require any previous structural assumption. The applicability of the method is therefore very general, and could be employed both in atomically-resolved or lower resolution images. However, the quality of the final result will strongly depend on the capability of reconstructing images with a level of detail that could grant a precise estimation of the sample drift.

\subsection{Reducing dose accumulation effects}

In recent years, the need of reducing radiation damage in STEM has promoted the development of new scanning modes. The acquisition of random sparse images combined with the capability of reconstructing the target image with adapted post-processing techniques has emerged as an efficient way to reduce total electron dose and dose accumulation effects.
A main drawback of the method is the need to know \textit{a priori} the sampling rate consistent with a limited loss of information.
In the random-scan mode, the complete image is recorded and therefore no image reconstruction is required. The random order of the pixel acquisition, and the wide distance between successively illuminated area, limits dose accumulation effects. To illustrate the advantage of the method we show the difference between sequential and random scan on nano-CL images of point defect-related emissions.

Hexagonal boron nitride ($h$-BN) is a wide band-gap semiconductor presenting strong excitonic emissions in the deep ultraviolet (UV), at $\sim$6 eV.\cite{watanabe_direct-bandgap_2004,cassabois_hexagonal_2016} Recently, it has been demonstrated that several defect-related emissions may act as very stable and bright single-photon emitters (SPE) in the visible and UV spectral range, opening the way to their potential usage in the quantum optics field.\cite{tran_quantum_2016,bourrellier_bright_2016,tran_robust_2016,jungwirth_temperature_2016,chejanovsky_structural_2016,shotan_photoinduced_2016,Martinez_efficient_2016,	exarhos_optical_2017,caldwell2019photonics}
The most studied centers present a narrow zero phonon line (ZPL) in the 1.7--2.3 eV energy range\cite{jungwirth_temperature_2016} followed by phonon sidebands red-shifted by about 65 meV and 160 meV. 
These SPEs have stable energies when excited by green laser but exhibit a
strong spectral diffusion (time-variable fluctuation in energy) under blue laser excitation.\cite{shotan_photoinduced_2016} Finally, a blinking behavior (intermittent emission) has also been described under prolonged illumination.\cite{Martinez_efficient_2016}

Nano-CL within a STEM is a very effective technique to investigate the spatial distribution of defect-related emissions at sub-wavelength resolution,\cite{Zagonel2010,KOCIAK2017112} and when combined with a Hanbury-Brown and Twiss interferometer, to test their potential non-classical nature.\cite{tizei_spatially_2013,meuret_photon_2015}
In a recent work, our group has shown that UV emitting centers in $h$-BN appear, when imaged by nano-CL, as highly localized spots whose diameter ($\sim$80 nm) corresponds to the characteristic excitation diffusion length of the material.\cite{bourrellier_bright_2016} The same methodology and experimental setup is applied here to investigate defect emissions in $h$-BN in the visible range.

\begin{figure}[tb]
	\centering
	\includegraphics[width=\columnwidth]{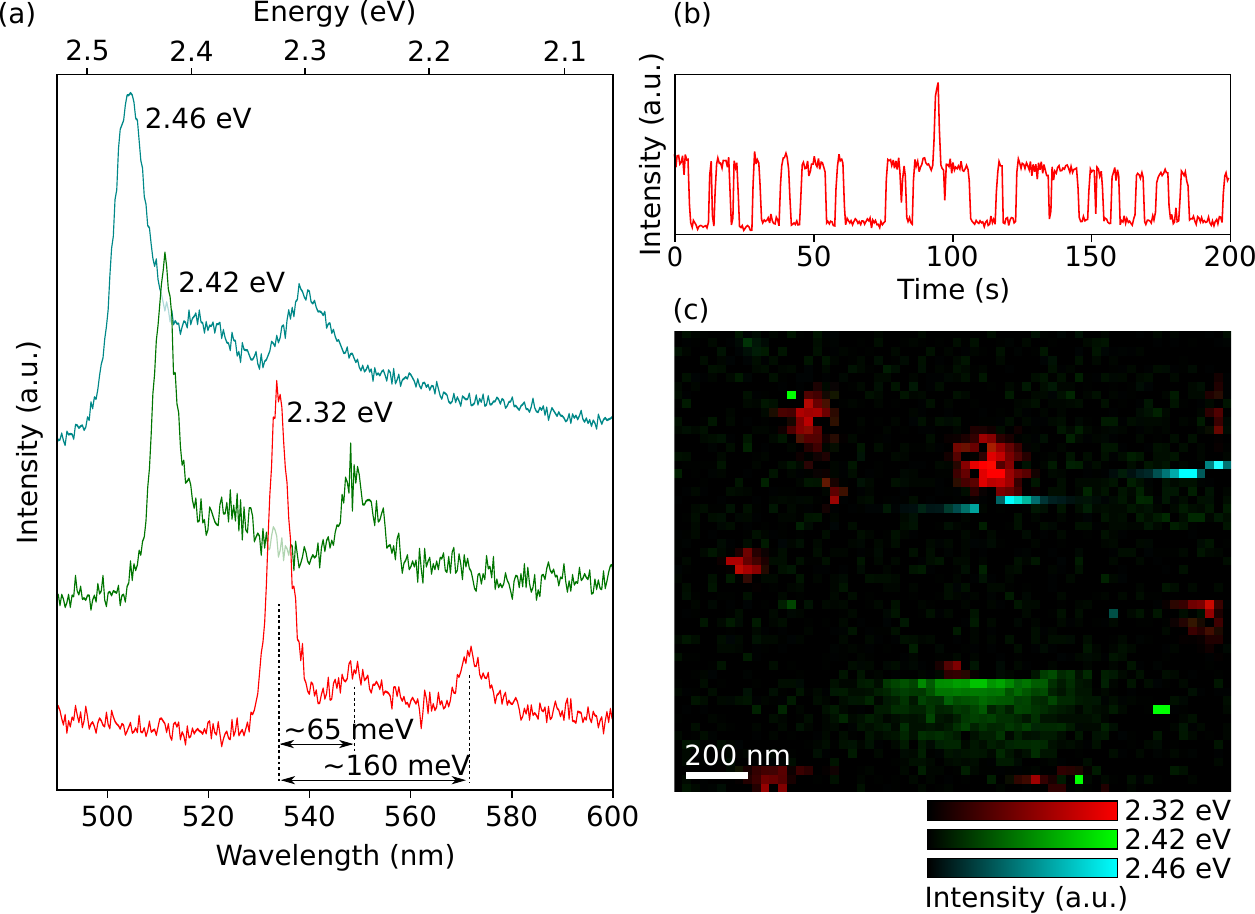}
	\caption{(a) CL of defects emissions in the visible range from an exfoliated $h$-BN flake. (b) Time evolution of the emission intensity under continuous illumination of an individual spot emitting at 2.32 eV. (c) Composite CL intensity map acquired in the same flake using the standard raster scan mode. Intensities have been obtained, after background subtraction, by integrating the signal in a 10 meV energy windows centered about the ZPL maxima. The colour code follows the schema of panel (a), each color scale has been normalized independently.\label{BN-visible-CL}}
\end{figure}

Fig. \ref{BN-visible-CL}(a) shows a few examples of CL spectra of a class of optically-active centers of $h$-BN on which their typical phonon replica structure of sidebands at 65 meV and 160 meV can be observed. With respect to photoluminescence typically carried out with a 532 nm (2.33 eV) green laser excitation, high-energy electrons employed in CL permit an easier access to higher energy excitations; here we report ZPLs in the 2.3--2.5 eV energy window which extends the energy range previously reported for this class of defects.
The time dependence of the emission intensity of the 2.32 eV line, obtained by continuous illumination of an individual emission spot of 30 nm $\times$ 30 nm in size is shown in Fig. \ref{BN-visible-CL}(b). We observe a blinking behavior in the tens of seconds time scale, in agreement with previous reports by photoluminescence.\cite{Martinez_efficient_2016} This effect has a strong impact when performing hyperspectral images in a raster scan mode, where time and space information are coupled. As an example, Fig. \ref{BN-visible-CL}(c) is a post-processed regular CL hyperspectral image acquired at an individual $h$-BN flake where a high density of different emitting centers from this class can be distinguished. CL intensity maps have been derived for each emission line by integrating the ZPL with an energy window of $\sim$10 meV centered about the intensity maximum.
The 2.32 eV emissions (in red) appear as isotropic spots with a full-width at half-maximum of $\sim$100 nm. Within the emission spots, stripes of dark pixels can be clearly discernible, which correspond to off states of the blinking centers. Other emission centers can be suddenly activated or deactivated when the electron beam position gets close to the emitting defect. This effect can lead to emissions localized on just a few consecutive pixels or emission stripes on sole two consecutive scanning lines, as clearly seen for the 2.42 eV and 2.46 eV centers in green and cyan, respectively. These intensity instabilities are likely to be driven by the electron beam, and may involve a variation in the charge state of the defect centers: charge can progressively accumulate due to secondary electrons ejection, and it can drive the system to an on/off state, which can switch after a sudden discharge. The charge hypothesis for time instabilities has been previously suggested in Ref. \citenum{Martinez_efficient_2016}, where charge variations were optically-induced.

\begin{figure*}[tb]
	\centering
	\includegraphics[width=0.7\textwidth]{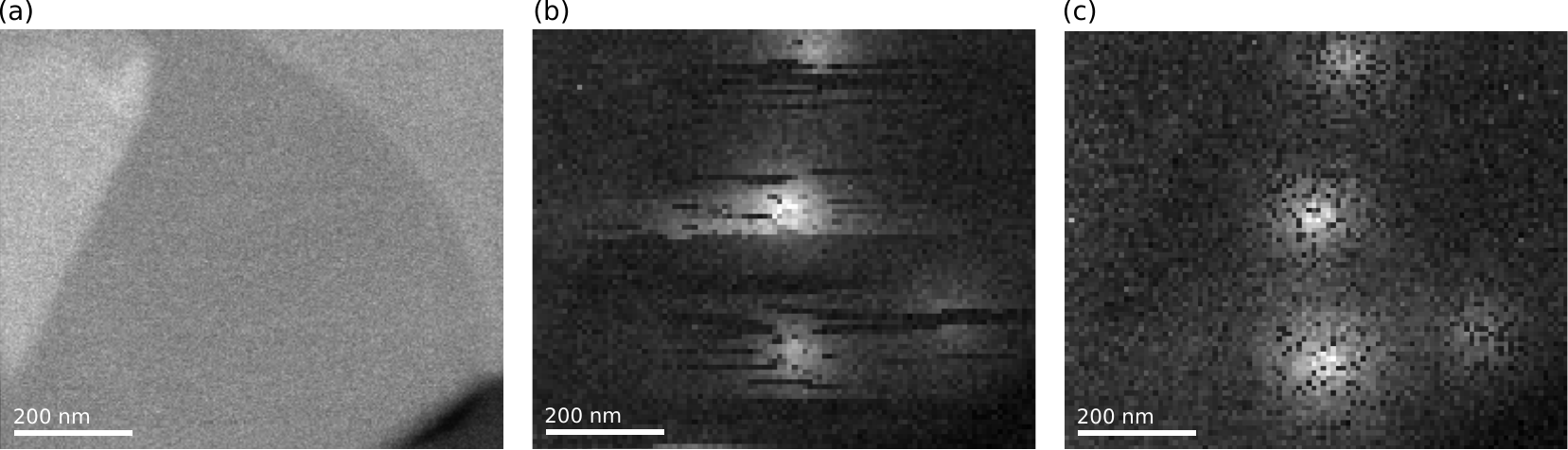}
	\caption{(a) HAADF image of an individual $h$-BN flake. (b) Corresponding CL intensity map of 2.32 eV emission centers obtained using the standard sequential scanning mode, and (c) using the new random scan mode. \label{randomvsraster}}
\end{figure*}

Emission instabilities strongly degrade the spatial information that can be obtained from hyperspectral images, but considerable improvements can be achieved by operating in the random scan mode. Fig. \ref{randomvsraster}(a) shows the HAADF image of a chemically-exfoliated $h$-BN flake. Fig. \ref{randomvsraster}(b) presents the CL intensity map of the 2.32 eV center extracted from a nano-CL hyperspectral-image acquired in standard raster mode, and in Fig.\ref{randomvsraster}(c), the map of an equivalent area obtained in random scan mode. In both intensity maps, four well-localized emitting centers can be identified but, in the raster scan mode extended dark stripes and abrupt intensity variation are observed; whereas in the random scan mode, emission spots are well-defined and only few dark pixels appear. In the sequential scan mode, adjacent pixels are consecutive in time and charges can accumulate leading to instabilities in the emission map, similar to artifacts observed in SEM images of insulating samples. In the random scanned emission map, this effect is strongly hindered: the average distance between successive pixels is 414 nm, while it is only 8.5 nm, corresponding to the pixel size, in the sequential scan.
It is reasonable to expect that the reduction of cumulative effects achieved with the random scan mode could provide significant improvements also in the spectroscopy of electron-beam sensitive materials which might be amorphized or even be destroyed at high electron doses.

\subsection{Time evolution in hyperspectral images}

In hyperspectral imaging, the observation of the temporal evolution of a spectroscopic signal with a spatial resolution is precluded by the intrinsic coupling of the space and time coordinates given the conventional sequential scanning. However, time-dependent changes can be considerably recovered by operating in the random scan mode, and employing a multi-step and iterative post-processing. Time-slicing and spatial integration permit a precise identification of even faint and non-recurrent spectroscopic features changing in time, and avoid spectral mixing. Once relevant signals are identified, their evolution with time can be monitored, the time windows over which spectroscopic features are stable can be determined, and corresponding time-resolved spectroscopic maps can be extracted. 

\begin{figure*}[tb]
	\centering
	\includegraphics[width=\textwidth]{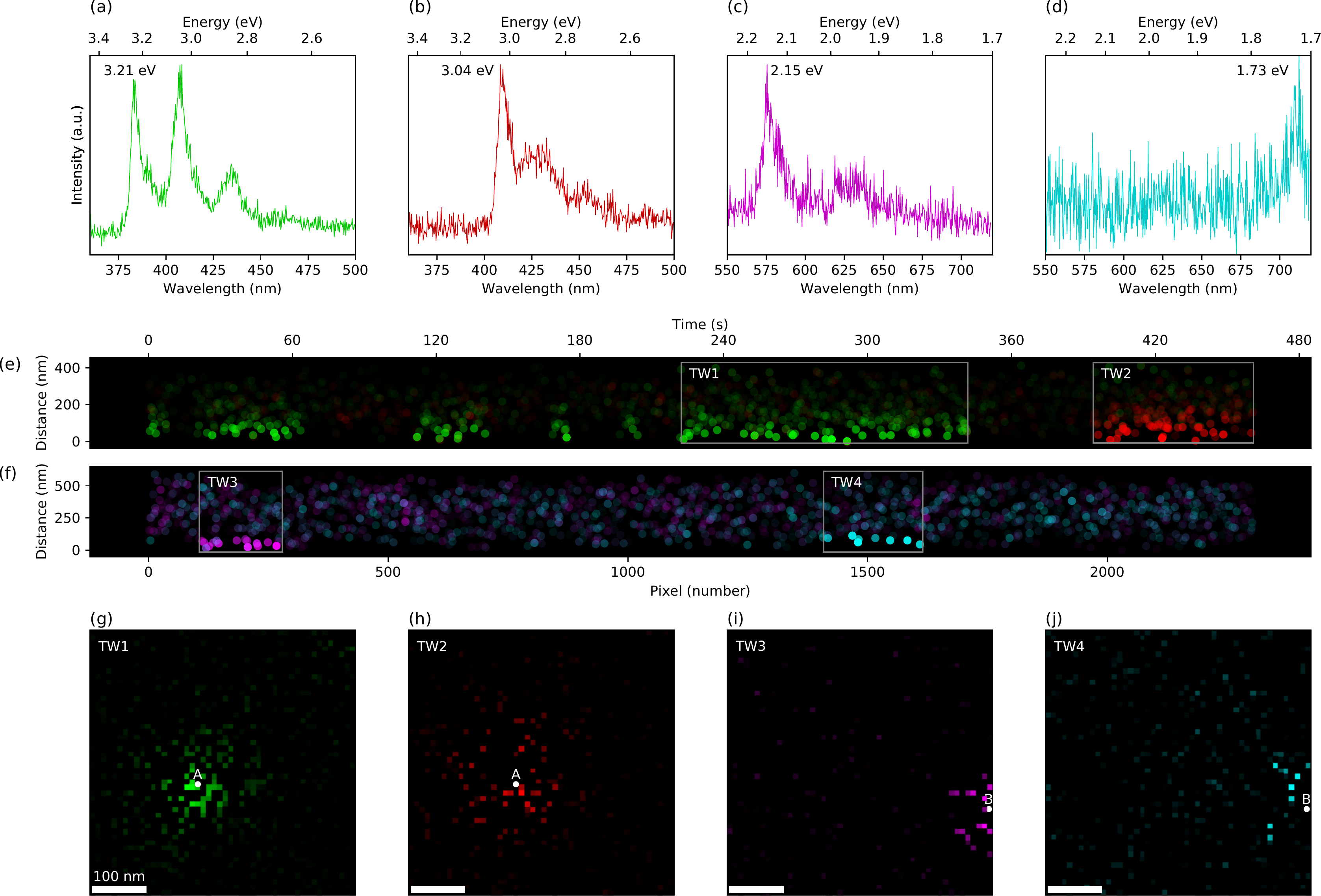}
	\caption{Space and time dependence of various CL signals in a few-layer $h$-BN flake obtained from a random-scan hyperspectral image, illustrating the blinking and spectral diffusion of individual emission spots. (a--d) Characteristic CL spectral features observed in this single hyperspectral image. (e,f) Time evolution of the emission lines as a function of the pixel acquisition time and pixel distance from the center of the emission origin. Emission intensities are encoded in the dots transparency. (g--j) Intensity maps extracted for the characteristic signals in (a--d) within the time windows (TW) indicated in panels (e,f). The points A and B indicate the reference positions used for the graphs in panels (e,f). \label{BN-random-scan}}
\end{figure*}

As discussed in the previous section, $h$-BN emitting defect centers present time instabilities, spectral diffusion and a highly localized CL signal, therefore they may serve as an ideal system to illustrate the potential of random scan to recover the time evolution in hypersectral images. Additional plasma treatment of $h$-BN-loaded TEM grids was used to generate a higher density of defect centers within the flakes.
Fig. \ref{BN-random-scan} presents the space and time dependence of several CL signals in few-layer $h$-BN obtained from a random-scan hyperspectral image acquisition. CL emission spectra extracted from appropriate time slices and spatial integration are given in Fig. \ref{BN-random-scan}(a--d); ZPLs followed by characteristic phonon replica can be observed at the 3.21 eV (386 nm), 3.04 (407 nm), 2.15 eV (576 nm) and 1.73 eV (714 nm) photon energies.
The two highest energy emissions [Fig. \ref{BN-random-scan}(a,b)] present a spectral overlap and therefore intensity maps cannot be extracted through simple signal integration within an energy window. Thus, we have employed a fitting routine using the spectra in Fig. \ref{BN-random-scan}(a--d) as reference functions. Emission intensity maps have been derived from the fitting coefficients with each map normalized independently. Appropriate slices of the maps for time windows over which the spectroscopic signal are active are provided in Fig. \ref{BN-random-scan}(g--j); spectra in Fig. \ref{BN-random-scan}(a--d) had been obtained by a further spatial integration of these slices.
Maps of Fig. \ref{BN-random-scan}(g,h) indicate that the 3.21 eV and 3.04 eV emission lines are centered in the same vicinity (point A).
An analysis of the time evolution of these spectral features is given in Fig. \ref{BN-random-scan}(e), as assessed by decomposing the full intensity maps individually and tracking the pixels based on the distance relative to an assigned origin (point A) as a function of the pixel acquisition time.
The dot brightness encodes the intensity of the emissions.
The minimal deviation in distance of the brightest dots over time for each of the emission lines implies minimal drift throughout the duration of the data acquisition. Additionally, the decay of the dot intensity variation with distance also corresponds to the excitation diffusion length ($\sim$90 nm) characteristic to such defects in $h$-BN. 
The gray boxes indicate the integrated time windows [TW1 and TW2] corresponding to Fig. \ref{BN-random-scan}(g,h), within which the spectroscopic signal are active. The alternation of dark and bright time frames is indicative of the blinking behavior of the intermittent emission center. Furthermore, the 3.21 eV and 3.04 eV emissions appear in well-separated time windows despite originating from the same vicinity; in no one pixel is a superposition of the two signals found. This observation is a clear indication of spectral diffusion at a single emitting spot, an effect which might be associated with a different charge state or with a structural degradation.
The 3.21 eV line [Fig. \ref{BN-random-scan}(a)] corresponds to a class of blue defect luminescence in $h$-BN;\cite{Korsaks2012,Berzina2016} the 3.04 eV line [Fig. \ref{BN-random-scan}(b)] is likely of similar luminescence origin but after the irreversible change in charge state or degradation with continued irradiation.

A similar analysis was performed for the 2.15 eV and 1.73 eV emission lines from Fig. \ref{BN-random-scan}(c,d). These spectral features occur only for a limited time frame [TW3 and TW4 in Fig. \ref{BN-random-scan}(f)] and the centers appear at the border of the scanned region [Fig. \ref{BN-random-scan}(i,j)]. Therefore the corresponding maps are less defined and it is more difficult to identify the emission center [point B has been used as reference origin for Fig. \ref{BN-random-scan}(f)]. However, we observe a spatial overlap between the emitting regions and a temporal separation, which indicate once more a possible spectral diffusion of the same emission center.
Such emission lines can be generated by way of high-temperature annealing in Ar or O$_2$ atmospheres, high-energy electron irradiation, such as in a SEM, ion irradiation or chemical etching.\cite{tran_robust_2016,chejanovsky_structural_2016} The color diversity within the same class of quantum emitter,\cite{jungwirth_temperature_2016,chejanovsky_structural_2016} photo-induced spectral diffusion, and blinking, suggest a change in the charge state of the emitter or the local crystal symmetry in the $h$-BN.\cite{Martinez_efficient_2016,shotan_photoinduced_2016} All of which stimulate interest as to how individual defect centers and their local environment can differ at the atomic-scale, and how that plays into the dynamics of their luminescence. The use of random scan mode with nano-CL is perfectly suited to investigate the temporal evolution of such phenomena at the nanoscale.

\section{Conclusions}

In this work, we have presented the implementation of a random scan operating mode in STEM obtained at the hardware level via a custom scan control module, and provided examples of its use in EELS and nano-CL hyperspectral imaging. With respect to the conventional raster scanning mode, probe displacements are longer and thus random scan might not be as suitable for simple STEM imaging, but this is not problematic in STEM spectro-microscopy for which longer illumination times are required. The reduction of dose-accumulation effects, and the possibility of decoupling the space and time information are significant advances brought on by the use of random scan to STEM hyperspectral imaging.
In the current implementation, the complete scan pattern is defined prior to the acquisition and then a full image matrix is recorded together with the pixel scanning order. Hyperspectral images can be filtered as a function of the pixel acquisition history, and image reconstruction techniques can be applied to these time series of subsampled random sparse images. By employing adapted post-processing tools, it is shown that the random scan method allows to track and correct for drift caused by sample instabilities, and to follow spectral diffusion with a high spatial localization. 

The practical implementation of the method permits a high control and flexibility of the acquisition parameters, and other operating modes can be later explored. Therefore, we foresee that STEM random scan could be generalized to a wide class of problems for which further specific protocols and post-processing tools might be developed. For instance, a constrain on the minimal distance between successive pixels could be imposed to further limit accumulation effects.
The technique can also be easily extended to feature-adapted dynamic sampling, which has been recently discussed as a way to combine high-resolution and low electron dose.\cite{dahmen2016feature,hujsak2018,stevens2018sub}
Indeed, at the current stage, reconstructed images are already generated live in real-time from subsampled data sets; this information can be employed to re-actualize the scanning path during the acquisition. Progress in image and hyperspectral image reconstruction will also benefit from key inputs and validator tests obtained from a direct recording of fully-random sparse images. Furthermore, the possibility to elaborate live partial dataset will motivate the development of faster algorithms of high accuracy and the interlink between microscopes and modern computing architectures. Finally, improvements in the microscope scan devices at the hardware level that reduce displacement times to the $\mu$s range, or even shorter, will extend the interest and applicability of random scan mode also to STEM imaging.

\appendix*
\section{Methods}

A \ce{NdNiO3} (NNO) thin film was grown by pulsed laser deposition  over a \ce{LaAlO3} (LAO) substrate and a cross-section lamella for electron microscopy was prepared along the [100] pseudocubic zone-axis of the substrate by focused ion beam (FIB) using a FEI Scios DualBeam. More details on the interface growth and specimen preparation can be found in Ref. \citenum{Preziosi-18}. Atomically-resolved EELS mapping was performed on a Nion UltraSTEM 200 microscope operated at 100 keV.

Few-layer \textit{h}-BN flakes have been obtained by liquid-phase exfoliation of commercially available micrometric powder following the protocol presented in Ref. \citenum{Coleman568}. A solution of $h$-BN powder in isopropanol (1 mg/mL) was sonicated for 1--10 h, and then drop-casted onto a lacey carbon TEM grid. Flakes presented a lateral size of few microns and an average thickness of few tens of nm. Additional plasma charging of the $h$-BN-loaded TEM grid in 25\% O$_2$/Ar atmosphere was used to generate/activate more defect centers for the results presented in Fig. \ref{BN-random-scan}.
Nano-CL experiments were performed on a VG-HB501 dedicated STEM operated at 60 keV, below the atomic displacement threshold for \textit{h}-BN in order to limit radiation damage. The microscope was provided with a liquid nitrogen cooling system for the sample stage (150 K). 
CL signals were collected using an Attolight M\"onch 4107 STEM-CL system fitted with an optical spectrometer with a 150-grooves/mm diffraction grating blazed at 500 nm. The spectral resolution of the spectrometer CCD was 0.34 nm/pixel. Hyperspectral images were obtained by recording one full CL spectrum per pixel while scanning the sample. Typical dwell times were 100--300 ms per pixel.

\acknowledgments{
We acknowledge support from the Agence Nationale de la Recherche 
(ANR), program of future investment TEMPOS-CHROMATEM and TEMPOS-NANOTEM (No. ANR-10-EQPX-50). The work has also received funding from the European Union in the Horizon-2020 Framework Programme (H2020-EU) under Grant Agreement No. 823717 (ESTEEM3). Authors would like to acknowledge Daniele Preziosi for the LAO-NNO thin film growth and Alexandre Gloter for the FIB lamella preparation.
}

\bibliography{biblio}

\end{document}